\documentclass[reprint,amsmath,amssymb,aip]{revtex4-1}
\usepackage{graphicx}
\usepackage{dcolumn}
\usepackage{bm}
\usepackage[]{natbib}
\usepackage{subfigure}
\usepackage{color}
\DeclareMathOperator{\sgn}{sgn}
\makeatletter
\newcommand*\xbar[1]{%
  \hbox{%
    \vbox{%
      \hrule height 0.5pt 
      \kern0.5ex
      \hbox{%
        \kern-0.1em
        \ensuremath{#1}%
        \kern-0.1em
      }%
    }%
  }%
} 
\begin{document}
\preprint{Nonlinear Landau damping}
\title{Landau damping of electron-acoustic waves  due to multi-plasmon resonances}
\author{Amar P. Misra}
\email{apmisra@visva-bharati.ac.in; apmisra@gmail.com}
\affiliation{Department of Mathematics, Siksha Bhavana, Visva-Bharati University, Santiniketan-731 235,  India}
\author{Debjani Chatterjee}
\email{chatterjee.debjani10@gmail.com}
\affiliation{Department of Applied Mathematics, University of Calcutta,  Kolkata-700 009, India}
\author{Gert Brodin}
\email{gert.brodin@physics.umu.se}
\affiliation{Department of Physics, Ume{\aa} University, SE-901 87 Ume{\aa}, Sweden}
\begin{abstract}
The linear and  nonlinear theories of electron-acoustic waves (EAWs) are studied in a partially degenerate quantum plasma with two-temperature electrons and stationary ions.  The initial equilibrium of electrons is assumed to be given by the  Fermi-Dirac distribution  at finite temperature.  By employing the  multi-scale asymptotic expansion technique to the one-dimensional Wigner-Moyal and Poisson equations, it is shown that the effects of  multi-plasmon resonances lead to a modified complex Korteweg-de Vries (KdV) equation with  a new nonlocal nonlinearity.  Besides giving rise to a  nonlocal nonlinear term, the wave-particle resonance  also modifies the local nonlinear coupling coefficient of the KdV equation. The latter is shown to conserve the number of particles, however, the wave energy decays with time. A careful analysis shows that the two-plasmon resonance is the dominant  mechanism for nonlinear Landau damping of EAWs.  An approximate soliton solution of the  KdV equation is also obtained,  and it is shown that the nonlinear  Landau damping causes the wave amplitude to decay   slowly with time compared to  the classical theory. 

\end{abstract}
\maketitle
\section{Introduction}\label{sec-intro} In many equilibrium plasmas,  electrons can be divided into two  groups with different temperatures \cite{goswami1976}. Such a division may be possible as certain electrons are preferentially heated by some external sources like beams or waves until a final equilibrium is reached.  The existence of two-temperature electrons was also experimentally observed in a sputtering magnetron plasma \cite{sheridan1991}. However, in $1977$, it was investigated by Watanabe and Taniuti \cite{watanabe1977} that if electrons in a plasma with background ions have  two different temperatures, there can exist an electron-acoustic wave (EAW). They also reported that if the hot electron number density is much higher than that of cold ones, the phase velocity of such waves lies in between the thermal velocities of cold and hot species, and the EAWs do not experience strong Landau damping. Later, in $1991$, Holloway and Dorning \cite{holloway1991} noted that a small-amplitude undamped EAW can exist in plasmas with the dispersion relation of the form $\omega\approx 1.31k v_\text{th}$, where $\omega~(k)$ is the wave frequency (number) and $v_{th}$ is the electron thermal speed.   Recently,  based on the observations from the Van Allen Probes, Vasko \textit{et al.} \cite{vasko2017} reported   the existence of EAWs  in the inner magnetosphere with the similar properties as above but a different linear dispersion law (as per their notations): $\omega=kv_0\left(1-k^2d^2/2\right)$, where $v_0$ and $d$ are, respectively, the electron-acoustic velocity and dispersive scale. They also elucidated    the observed asymmetric solitary waves in terms of   electron-acoustic solitons and  shocks. In an another work by    Dillard \textit{et al.} \cite{dillard2018} it has   been shown by means of theoretical and numerical analysis of hydrodynamic equations that the asymmetric solitary waves can appear due to the steepening of initially quasi-monochromatic EAWs. The beam-generated EAWs  due to trapping of electrons by whistler waves has also been reported with the phase velocity slightly large than that of the whistler waves \cite{an2019}. Although, such EAWs have similar  properties with those stated before, their phase velocity is predicted a bit higher, i.e., $1.8\lesssim \omega/kv_\text{th}\lesssim2.2$.     
\par 
The EAWs are typically electrostatic plasma oscillations which propagate with a frequency below that of the Langmuir wave. The characteristics of EAWs have been extensively studied theoretically  by means of Vlasov-Poisson simulation \cite{holloway1991,valentini2012} and PIC simulation \cite{valentini2006},  as well as experimentally \cite{anderegg2009,chowdhury2017}.  However, all these investigations were limited to the classical regimes. Recently, a liner theory of EAWs in degenerate plasmas with two groups of electrons has been studied by Fahad \textit{et al.} \cite{fahad2019}. They considered a fully degenerate plasma with two different Fermi temperatures of electrons. However, such a model  may not be physically sound in the context of quantum mechanics  as there should not be more than one species of electrons in a fully degenerate plasma or a  zero-temperature Fermi gas. 
\par In plasmas with two-temperature electrons, the nonlinear theory of EAWs has been investigated by many authors, however, using the fluid theory approach, i.e., without the Landau damping effects \cite{lakhina2008,akter2016}. On the other hand, after the work of Ott and Sudan \cite{ott1969} on nonlinear theory of ion-acoustic waves with linear Landau damping, many authors have studied the Landau damping of low-frequency electrostatic waves \cite{vandam1973,misra2015,barman2017} as well as high-frequency wave envelopes \cite{chatterjee2015,misra2017}, however, no nonlinear  theory of EAWs with Landau damping has  been developed in the context of quantum plasmas.   While the linear plasmon resonance is known in the literature, the multi-plasmon resonance \cite{brodin2017} in the nonlinear regime has not yet been fully explored, especially in the context of Korteweg-de Vries (KdV) equation. 
\par 
In this work, we will generalize the previous works on EAWs to include short-scale quantum effects. This is of interest for dense plasmas  where the characteristic de Broglie wavelength of electrons is larger than the nearest neighboring distance. Such dense plasmas include those appear in various astrophysical environments and certain laboratory systems, e.g., cases where high power lasers compress solid density targets. We consider weakly coupled systems only  such that a background distribution, deviating from the thermodynamic equilibrium, can be sustained. More specifically, we study the linear and nonlinear theories of EAWs in a partially degenerate plasma consisting of two groups of electrons with their thermodynamic temperatures $(T_j)$ exceeding the Fermi temperature $(T_{Fj})$ somewhat (i.e., $T_j>T_{Fj}$, where $j=l,h$ stand for low and high temperature electrons) and stationary positive ions. The purpose of this study is to find the quantum generalization of known classical results of EAWs including wave-particle interaction. As a consequence, we need to consider the Wigner theory rather than the classical Vlasov or semi-classical Vlasov theories to take into account the effects of quantum dispersion and the wave-particle resonance. A specific and novel feature introduced by the Wigner theory, is a nonlinear wave-particle interaction referred to as multi-plasmon interaction (see e.g. Refs. \citep{misra2017,brodin2017}) where we can have simultaneous absorption (or emission) of multiple plasmon quanta. Including this mechanism into the calculation scheme, an additional wave damping due to the two-plasmon resonance is seen to occur in the evolution of EAWs  along with the resonance of linear theory. For these types of effect to be significant, we have to focus on the regime of short-wavelengths, i.e., the thermal de Broglie wavelength can not be much smaller than that of EAWs $(\hbar k/m\lesssim v_{tl})$ which still accounts for some quantum effects to enter the picture. Here, $\hbar=h/2\pi$ is the reduced Planck's constant, $k$ is the wave number, $m$ is the electron mass, and $v_{tl}$ is the thermal velocity of low-temperature electrons.
\par
 An important aspect for the details of EAWs is the properties of the
electron background distribution function. For dense plasmas, as can be found, e.g.,  in the interiors of giant stars like white dwarfs and neutron stars \cite{horn1991,chabrier2002}, as well as in gas giants (e.g., Jupiter) \cite{coppari2013},  typically the density is high
enough for the electron distribution function to be degenerate or partially degenerate. In our study, the case of partial degeneracy is in the focus. Moreover, we are concerned with the background distributions deviating from the thermodynamic equilibrium. The high density
(partially degenerate) plasmas where the electron background deviates from the thermodynamic equilibrium can appear in the context of plasmas produced by lasers or ion beams \cite{hau-riege2011,gibbon2005}. Importantly, the system energy flows mostly into the electrons, and, as a result, there may appear partially degenerate electrons with high temperature tails, allowing the
electron background to be modeled by two different temperatures. In such systems, EAWs are likely to be excited and play a role in the nonlinear dynamics. 
\par   The paper is organized as follows:  In Sec. \ref{sec-model} we present the basic physical model   and  the underlying   assumptions.    The linear theory of EAWs comprising the wave dispersion and Landau damping  are studied in Sec. \ref{sec-linear}. Section \ref{sec-derivation} presents the derivation of the modified KdV equation and the occurrence of multi-plasmon resonances. We have verified the conservation laws for the  KdV equation as demonstrated in Sec. \ref{sec-conser}. In Sec. \ref{sec-soliton}, an approximate soliton solution of the KdV equation is   obtained and the effect of Landau resonance is discussed. Finally, Sec. \ref{sec-conclusion} is left to summarize  and conclude the results.         
\section{The model}\label{sec-model} We consider the nonlinear wave-particle interaction and evolution of small-amplitude  EAWs in a  partially degenerate plasma with two-temperature electrons and stationary ions. Our basic assumptions for the study are the following:
\begin{itemize}
\item Under certain conditions electrons  have a relatively high-temperature tail such that electrons can be grouped into two distinct components with different thermodynamic temperatures $T_l$ and $T_h$ [Hereafter we call  low- $(\alpha=l)$ and high-temperature $(\alpha=h)$ electrons with $T_h>T_l$ and $T_\alpha \gtrsim T_{F\alpha}$, where $T_{F\alpha}$ denotes the    Fermi temperature of $\alpha$-species electrons].
\item Although, the theory is independent of the background distribution,  however,  we consider the initial equilibrium of   electrons as given by the Fermi-Dirac distribution at finite temperature $(T_\alpha\neq0)$. The case of full  degeneracy of electrons at zero temperature restricts only one species of electrons and thereby inadmissible to the present study.     
\item There exists a low-frequency mode with the frequency $(\omega)$ proportional to the wave number $(k)$ such that the linear wave damping is small and the nonlinear evolution of EAWs can be described by the Korteweg de Vries (KdV) equation. 
\item Similar to classical plasmas, the equilibrium number density of high-temperature electrons is higher than that of low-temperature species, i.e.,  $n_{h0}> n_{l0}$. 
\item The EAWs are weakly dispersive such that  in the regime $\hbar k/m\lesssim v_{tl}$,   the Wigner-Moyal equation is still valid and some quantum effects come into play in  the process of wave-particle interactions.    

\end{itemize}   
We start with the   Wigner-Moyal equation, given by,
\begin{equation}
\begin{split}
\frac{\partial f_\alpha}{\partial t}&+{\bf v}\cdot\nabla_{\bf r} f_\alpha+\frac{iem^3}{(2\pi)^3\hbar^4}\int\int d^3{\bf r}' d^3{\bf v}' e^{im({\bf v}-{\bf v}')\cdot {\bf r}'/\hbar}\\
&\times \left[\phi\left({\bf r}+\frac{{\bf r}'}{2},t\right)-\phi\left({\bf r}-\frac{{\bf r}'}{2},t\right)\right]f_\alpha({\bf r},{\bf v}', t)=0, \label{eq-wigner}
\end{split}
\end{equation}
and the Poisson equation
\begin{equation}
 \nabla^2 \phi=4\pi e \sum_{\alpha=l,h} \left(\int f_\alpha d^3v-n_{0}\right), \label{eq-poisson}
\end{equation} 
where $f_\alpha$ is the Wigner distribution function for $\alpha$-species electrons; $e,m,v$, respectively, denote the charge, mass, and velocity of electrons, $\phi$ is the self-consistent electrostatic potential, and    $n_0$ is the number density of background ions. Moreover, we consider the   propagation of EAWs along the $x$-axis and the background   distribution function for $\alpha$-species electrons in one-dimension as   (writing $v_x$ as $v$) \cite{manfredi2005}
 \begin{equation} 
\begin{split}
 f_\alpha^{(0)}(v)&=\int\int f^{3D}_\alpha({\bf v})dv_ydv_z\\
 &=2\left(\frac{m}{2\pi\hbar}\right)^3\int\int\left[1+\exp\left(\frac{E-\mu_\alpha}{k_BT_{\alpha}}\right)\right]^{-1}dv_ydv_z\\
 &=\frac{3}{4} \frac{n_{\alpha0}}{v_{F\alpha}}\frac{T_\alpha}{T_{F\alpha}}\log\left[1+\exp\left(-\frac{\frac{1}{2}mv^2-\mu_\alpha}{k_BT_{\alpha}}\right)\right], \label{eq-distb-fn}
  \end{split}
\end{equation}
where $E=(m/2)(v_x^2+v_y^2+v_z^2)$ is the kinetic energy and  $\mu_\alpha$  the chemical potential which satisfies the following charge neutrality condition.
\begin{equation} \label{eq-mu}
n_0=\sum_\alpha\int f_\alpha^{(0)}(v) dv. 
\end{equation}
In the nondegenerate limit $(T_\alpha\gg T_{F\alpha})$, the parameter $\xi_\alpha=\mu_\alpha/k_BT_\alpha$ is large and negative, while it is large and positive in the fully degenerate limit $(T_\alpha\ll T_{F\alpha})$. We, however, consider the case of $T_\alpha\gtrsim T_{F\alpha}$ such that $\xi_\alpha~(<0)$ is of moderate value. 
\par
We stress here that there are certain parameter restrictions imposed by the Pauli-principle, since we cannot have a phase space density
of the background Wigner-function exceeding $2(m/2\pi\hbar)^{3}$. As a result, the parameters of the high- and low-temperature  distributions
cannot be chosen independently. The strictest criterion appears for $E=0$ and reads
\begin{equation}
\left[{1+\exp \left(-\frac{\mu _{l}}{k_{B}T_{l}}\right)}\right]^{-1}+\left[{1+\exp \left(-\frac{\mu
_{h}}{k_{B}T_{h}}\right)}\right]^{-1}\leq 1.
\end{equation}
For a partially degenerate low-temperature distribution (i.e., $k_{B}T_{l}\sim E_{F}$, $\mu _{l}\sim 2E_{F}$), this condition is typically fulfilled if the high-temperature distribution is not too far from the 
classical Maxwell-Boltzmann regime such that $-\mu _{h}/k_{B}T_{h}\gtrsim 3$. 
\section{Linear theory: wave dispersion and one-plasmon resonance} \label{sec-linear}
Before we proceed to the nonlinear theory of wave-particle interactions, it is imperative to study the linear theory of EAWs and the associated wave damping. To this end, we linearize  Eqs. \eqref{eq-wigner} and \eqref{eq-poisson} by splitting up the functions $f$ and $\phi$ into their equilibrium and perturbation parts, i.e., $f_\alpha(x,v,t)= f_\alpha^{(0)} +f_\alpha^{(1)}$ and 
$\phi(x,t)=\phi^{(1)}$, and assume the perturbations to be of the form $\sim\exp(ikx-i\omega t)$, i.e., plane waves with frequency $\omega$ and wave number $k$. Thus, we obtain the following dispersion relation.
\begin{equation}
D(\omega,k)\equiv 1-\sum_{\alpha=l,h}\frac{\omega_{p\alpha}^2}{n_{\alpha_0}k^2}\int_{-\infty}^{\infty} \frac{f_\alpha^{(0)}(v)}{(v-\omega/k)^2-{v_q^2}}dv=0, \label{eq-disp}  
\end{equation}
where   $\omega_{p\alpha}=\sqrt{4\pi n_{\alpha0}e^2/m}$ is the plasma frequency for $\alpha$-species electrons  and $v_q=\hbar k/2m$ is the velocity associated with the plasmon quanta.
\par
For weak wave damping of EAWs, the time asymptotic solution for the complex frequency $(\omega=\omega_r+i\gamma_L)$  can be obtained by finding  the roots of the dispersion equation $D(\omega,k)\equiv D_r(\omega_r,k)+iD_i(\omega_r,k)+i\gamma_L[\partial D_r(\omega_r,k)/\partial \omega_r]=0$. Thus, separating the real and imaginary parts we obtain 
 \begin{equation}
D_r(\omega_r,k)\equiv 1-\sum_{\alpha=l,h}\frac{\omega_{p\alpha}^2}{n_{\alpha0}k^2}{\cal P}\int \frac{f_\alpha^{(0)}(v)}{(v-\omega_r/k)^2-v_q^2}dv=0, \label{eq-Dr}  
\end{equation}
 and the Landau damping rate, given by,
 \begin{equation}
 \gamma_L=-\frac{D_i(\omega_r,k)}{\partial D_r/\partial \omega_r}, \label{eq-damp}
 \end{equation}
 where
 \begin{equation}
 D_i= -2\pi v_q\sum_{\alpha=l,h}\frac{\omega_{p\alpha}^2}{n_{\alpha0}k^2}\left[ f_\alpha^{(0)}(v^\text{res}_{+1})-f_\alpha^{(0)}(v^\text{res}_{-1})  \right],  \label{eq-Di}
 \end{equation}
in which  $v^\text{res}_{\pm}=\omega_r/k\pm  v_q$  denotes the plasmon resonance velocity.  
\par
 Next, substituting the distribution function \eqref{eq-distb-fn} into Eq. \eqref{eq-Dr} and evaluating the integrals in two different regimes $|v|<\omega_r/k\pm v_q$ and $|v|>\omega_r/k\pm v_q$, i.e., 
 \begin{equation} 
 \begin{split}
{\cal P}\int_{-\infty}^{\infty}&=\lim_{\epsilon\rightarrow0+}\left[\int_{-\infty}^{-(\lambda\pm v_q)}+\int_{-(\lambda\pm v_q)}^{(\lambda\pm v_q)-\epsilon}+\int_{(\lambda\pm v_q)+\epsilon}^{\infty}\right]\\
&=\lim_{\epsilon\rightarrow0+}\left[\int_{-(\lambda\pm v_q)}^{(\lambda\pm v_q)-\epsilon}+2\int_{(\lambda\pm v_q)+\epsilon}^{\infty}\right], \label{eq-int}
\end{split}
\end{equation}  
and noting that the exponential function in the distribution function [Eq. \eqref{eq-distb-fn}] is small in the partially degenerate plasma regime, we obtain
 \begin{equation}
 \begin{split}
 D_r(\omega_r,k)\equiv 1-\frac{3}{4}&\sum_{\alpha=l,h}\omega_{p\alpha}^2e^{\xi_\alpha}\left[2\sqrt{2}\frac{v_{t\alpha}/v_{F\alpha}}{\omega_r^2-k^2v_q^2}\right.\\
 &\left.+\frac{\sqrt{2\pi}v_{t\alpha}-\omega_r/k}{k^2v_{t\alpha}^2v_{F\alpha}}   \right]=0. \label{eq-Dr1}
 \end{split}
 \end{equation}
Equation \eqref{eq-Dr1} describes both the high-frequency and relatively low-frequency wave eigenmodes of EAWs.  In the limit of $\omega_r/k\pm v_q\gg v_{th}$, the high-frequency Langmuir wave (LW) mode can be obtained from Eq. \eqref{eq-Dr1} by considering the first and the second terms as
\begin{equation}
\omega_r^2=k^2v_q^2+\frac{3}{\sqrt{2}}\sum_{\alpha=l,h}\omega_{p\alpha}^2\frac{v_{t\alpha}}{v_{F\alpha}}e^{\xi_\alpha}, \label{eq-LW}
\end{equation}
where $v_{t\alpha}=\sqrt{2k_BT_\alpha/m}$ is the thermal velocity and $v_{F\alpha}=\sqrt{2k_BT_{F\alpha}/m}$ the Fermi velocity of $\alpha$-species electrons.
\par
\begin{figure*}! 
\includegraphics[scale=0.5]{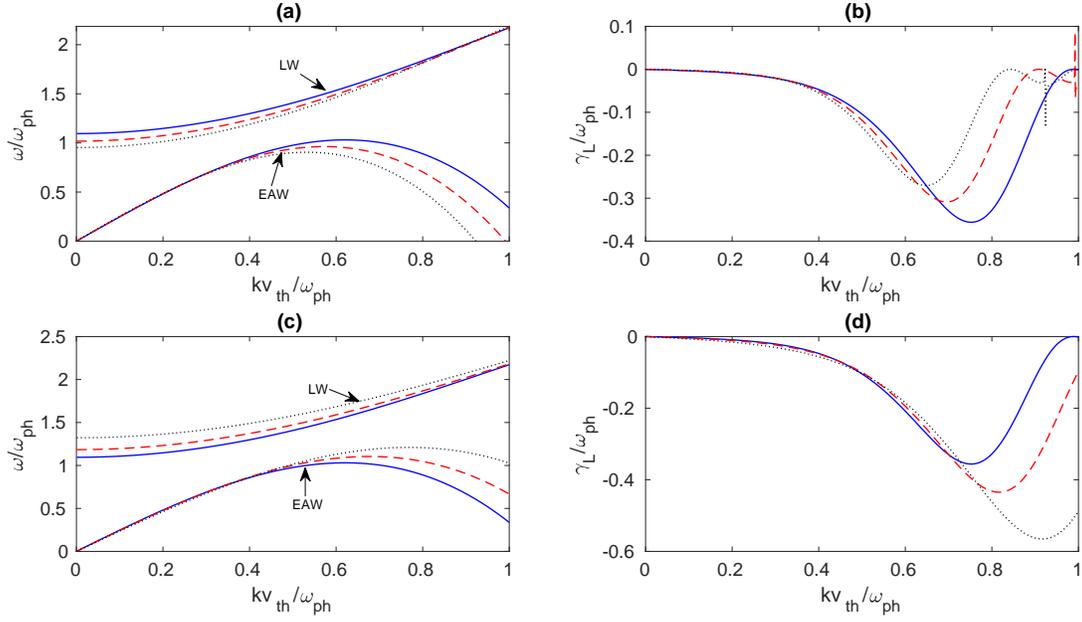}
\caption{The  dispersion curves   for   Langmuir waves (LWs) and electron-acoustic waves (EAWs) [Subplots (a) and (c)],   given by Eqs. \eqref{eq-LW} and \eqref{eq-EAW}, and the Landau damping rates for EAWs [Subplots (b) and (d)],   given by  Eq. \eqref{eq-damp},     are shown. For subplots (a) and (b), the fixed parameters are $n_{l0}=2\times10^{24}$ cm$^{-3}$,  $T_l=10^6$ K and $T_h=2.5 T_l$; the solid, dashed and dotted lines, respectively, correspond to  $n_{h0}=10n_{l0}$, $11n_{l0}$ and $12n_{l0}$.  For subplots (c) and (d), the fixed parameters are $n_{l0}=2\times10^{24}$ cm$^{-3}$, $n_{h0}=10n_{l0}$ and  $T_l=10^6$ K; the solid, dashed and dotted lines, respectively, correspond to  $T_h=2.5 T_l$, $2.7 T_l$ and $3 T_l$. }
\label{fig:disp-curve}
\end{figure*}

 
On the other hand, for $\omega_r/k\pm v_q\ll v_{tl}$, the first and the third terms of Eq.  \eqref{eq-Dr1} can be combined to yield the following dispersion relation for the   EAW mode.
\begin{equation}
\omega_r=k\left(\sqrt{2\pi}\sum_{\alpha=l,h}\frac{\omega_{p\alpha}^2e^{\xi\alpha}}{v_{t\alpha} v_{F\alpha}}-\frac{4}{3}k^2 \right)\Big/\sum_{\alpha=l,h}\frac{\omega_{p\alpha}^2e^{\xi\alpha}}{v^2_{t\alpha} v_{F\alpha}}. \label{eq-EAW}
\end{equation}
From Eq. \eqref{eq-EAW} we note that, similar to the ion-acoustic waves in an electron-ion plasma, the EAW in two-temperature-electron plasmas   has also the cubic order of dispersion and for small wave numbers  $\omega/k$ approaches a constant value, i.e., the EAW becomes dispersionless. These are the prerequisites for the description of small-amplitude EAWs by a KdV-like equation. 
Further inspecting on Eq. \eqref{eq-EAW} we find that   for typical plasma parameters $n_{l0}=2\times10^{24}$ cm$^{-3}$, $n_{h0}=10-12n_{l0}$,  $T_l=10^6$ K and $T_h=2.5-3 T_l$, and with $kv_{th}/\omega_{ph}\equiv k\lambda_h\ll1$,  the  ratio of the two terms (without the factor $k$ and the term involving $k^2$) on the right-hand side of Eq. \eqref{eq-EAW} roughly varies in between $2.2$ to $2.5$ times the thermal velocity of high-temperature  species ($v_{th}$). In this situation,    Eq. \eqref{eq-EAW} reduces to $\omega_r\approx 2.5kv_{th}$. Thus, it follows that the phase velocity of EAWs in a plasma with finite temperature degeneracy  is a bit higher than predicted in the classical theory \cite{holloway1991} $(\omega_r\approx 1.31kv_{th})$. Figure \ref{fig:disp-curve} displays the dispersion curves for high-frequency LWs and low-frequency EAWs [panels (a) and (c)], as well as the Landau damping rates for EAWs [panels (b) and (d)].   It is  seen that with lower   values of the temperature ratio $T=T_l/T_h$ or with higher values of the density ratio $N=n_{l0}/n_{h0}$,    a critical value $K_c$ of $K\equiv k\lambda_h$ exists,  beyond which the EAW frequency  can turn over, going to zero, and then assume negative values.   Such a nature of the wave frequency, which is unusual in classical plasmas where the high- and low-frequency branches join to form a thumb-like curve \cite{holloway1991}, may be due to finite temperature degeneracy of background electrons.    As a result, the linear damping rate $|\gamma_L/\omega_{ph}|$, which was initially low $(\sim0.01)$, starts increasing with certain values of  $K<K_c$ and then   decreases   at relatively higher values of $K>K_c$. This enables us to  consider the regime $0\lesssim K\lesssim 0.5$ or $\hbar k/mv_{th}\lesssim0.2$ where the linear Landau damping rate of EAWs remains weak.
\par 
 It is to be noted that the approximate dispersion Eqs. \eqref{eq-LW} and \eqref{eq-EAW} are obtained by assuming the smallness of the exponential function in the distribution function \eqref{eq-distb-fn} as applicable in the partially degenerate regime.  However, the Fermi-Dirac distribution approaches the Maxwell-Boltzmann distribution in the limit of low density or high temperature, i.e., when the integrand in Eq. \eqref{eq-distb-fn} is much smaller than the unity.     So, the classical results of LWs and EAWs cannot be recovered directly from Eqs.  \eqref{eq-LW} and \eqref{eq-EAW}.   
\par  
In what follows, we calculate the equilibrium chemical potentials   from the following relation. 
 \begin{equation}  
\sum_{\alpha=l,h}\int_{-\infty}^\infty f^{(0)}_{\alpha}(v)dv=n_0,  
\end{equation} 
 
which gives   
\begin{equation}
\frac{3}{4}\sqrt{\pi}\sum_{\alpha=l,h}\left(\frac{T_\alpha}{T_{F\alpha}}\right)^{3/2}n_{\alpha0}e^{\xi_\alpha}=n_0.
\end{equation}
 \section{Nonlinear theory: the KdV equation with multi-plasmon resonance}\label{sec-derivation}
On the basis of the results of the linear theory of EAWs (Sec. \ref{sec-linear}), i.e., EAW has a cubic order dispersion and the Landau damping rate is small, we proceed to the nonlinear theory of small-amplitude EAWs. Here, we derive an evolution equation for the weakly nonlinear EAWs in a degenerate plasma using the multiple-scale perturbation technique. Some special attention must be devoted to the higher order (in the amplitude) resonances that occur in the Wigner theory. 
In particular,
due to the nonlinearities, we will have Landau resonances with resonant
velocities that are shifted an amount $\pm n\delta v_{q}$ in momentum space
(See e.g., Ref. \citep{brodin2017}) compared to the resonant velocity of
classical theory (i.e., the phase velocity). Here, $n=1$ gives the velocity
shift  already appeared in the linear theory  as described in the previous
section \ref{sec-linear}. Close to the resonant velocities, the Wigner
equation must be analyzed in more detail. As a starting point, we will
divide the velocity space into resonant and non-resonant regions. While a
rigorous derivation of the nonlinear higher order resonances, based on the
Wigner equation, has been given in Ref. \citep{brodin2017}, we will here give a heuristic
derivation based on elementary quantum mechanics, following the steps of 
Ref. \citep{brodin2016} Our starting point is the quantum mechanical modifications of the
classical linear resonance, where the resonant velocity is adjusted
according to 
\begin{equation}
v^{\text{res}}=\frac{\omega }{k}\rightarrow v^{\text{res}}=\frac{\omega }{k}%
\pm \frac{\hbar k}{2m}.  \label{Eq-15}
\end{equation}%
Let us study the physical meaning of this modification. When a particle
absorbs or emits a wave quantum it can increase or decrease the momentum
according to 
\begin{equation}
\hbar k_{1}\pm \hbar k=\hbar k_{2},  \label{Eq-16}
\end{equation}%
and at the same time the energy changes according to%
\begin{equation}
\hbar \omega _{1}\pm \hbar \omega =\hbar \omega _{2}.  \label{Eq-17}
\end{equation}%
Next we identify $\hbar k_{1}/m$ (or equally well $\hbar k_{2}/m$) with the
resonant velocity $v^{\text{res}}$ and note that for small amplitude waves,
the particle frequencies and wavenumbers $(\omega _{1,2},k_{1,2})$ obey the
free particle dispersion relation $\omega _{1,2}=\hbar k_{1,2}^{2}/2m$.
Using these relations, we immediately see that the energy momentum relations
Eqs.~(\ref{Eq-16}) and (\ref{Eq-17}) imply the modification of the resonant
velocity seen in Eq.~(\ref{Eq-15}). An interesting possibility, which was
studied in Ref. \citep{brodin2017}, is the simultaneous absorption (or emission)
of multiple wave quanta, rather than a single wave quantum at a time. In
that case Eqs. (\ref{Eq-16}) and (\ref{Eq-17}) are replaced by 
\begin{equation}
\hbar k_{1}\pm n\hbar k=\hbar k_{2},  \label{Eq-18}
\end{equation}%
and 
\begin{equation}
\hbar \omega _{1}\pm n\hbar \omega =\hbar \omega _{2},  \label{Eq-19}
\end{equation}%
where $n=1,2,3,\ldots $ is an integer. Accordingly, performing the same
algebra as for the linear case, the resonant velocities now becomes%
\begin{equation}
v_{\pm n}^{\text{res}}=\frac{\omega }{k}\pm n\frac{\hbar k}{2m}.
\label{Eq-20}
\end{equation}%
When we pick the minus sign in Eq.~(\ref{Eq-20}), the resonant velocity for
absorbing multiple wave quanta can be considerably lower, provided the
wavelengths are short. As a consequence, in the case of Langmuir waves, the
damping rate due to absorption of multiple wave quanta can be much larger
than the standard linear damping rate. This is due to the larger number of
resonant particles in the former case.  
 
\par 
For the non-resonance region of velocity space we have $|v-v^\text{res}_{\pm n}|\gg o(\epsilon)$ with $\epsilon~(>0)$ denoting the scaling parameter and $v^\text{res}_{\pm n}=\omega/k\pm n\hbar k/2m$, $n=0,1,2,3,...$ the resonant velocity. 
 In the non-resonance regime, we consider the following expansions for $f_\alpha$ and $\phi$. 
\begin{equation}
\begin{split}
f_\alpha(x,v,t)=& f_\alpha^{(0)} +\epsilon^{1/2}f_\alpha^{(1)}+\epsilon^2 f_\alpha^{(2)}+\cdots,\\
\phi(x,t)=&\epsilon \phi^{(1)}+\epsilon^{3/2} \phi^{(2)}+\cdots\\   
\end{split} \label{eq-expan1}
\end{equation}
  However, in the resonance region where $|v-v^\text{res}_{\pm n}|\sim o(\epsilon)$,  a slightly different ordering  for both $f_\alpha$ and $\phi$ needs to be considered. This will be verified shortly.  The idea of considering different orderings for the distribution function and the electrostatic potential in the resonance and non-resonance regions was     proposed and rigorously discussed by Vandam and Tanuity \cite{vandam1973} on the consideration that the Landau damping of KdV solitons is a far-field approximation of the Vlasov equation. Although, the present nonlinear theory considers a similar approach, the mathematical treatment is a bit different from the above due to the Wigner function.  A particular ordering  for each of  $f_\alpha$ and $\phi$ cannot give rise the resonance    and nonresonance    contributions properly  rather with different orderings,  the terms involving the resonance and nonresonance  parts can be combined with the closure equation, i.e.,  the Poisson equation which involves both the resonance and nonresonance contributions of the distribution function.       
   Also,   the Gardner-Morikawa transformation, $\xi=\epsilon^{1/2}(x-\lambda t),~
\tau=\epsilon^{3/2} t$, which is applicable for the fluid theory  of a KdV equation, may not be directly applied to the Vlasov-Poisson or Wigner-Poisson systems because of the singularities in  the integrals due to resonant particles. Thus, we   modify the Gardner-Morikawa transformation as \cite{vandam1973,barman2017}
\begin{equation}
\xi=\epsilon^{1/2}x,~\sigma=\epsilon^{1/2}t,~\tau=\epsilon^{3/2} t    \label{eq-stretch}
\end{equation}
and assume  the following multi-scale  Fourier-Laplace transforms for the perturbations.
  \begin{equation}
 \begin{split}
  f_\alpha^{(n)}(v,\xi,\sigma,\tau)=& \frac {i}{(2\pi)^2} \int_W d\omega \int_ {-\infty}^{\infty}dk\tilde{f}_\alpha^{(n)}(v,k,\omega,\tau)\\
  &\times (\omega-\lambda k)^{-1} \exp[i(k\xi-\omega\sigma)] \\
 \phi^{(n)}(\xi,\sigma,\tau)= &\frac {i}{(2\pi)^2} \int_W d\omega \int_ {-\infty}^{\infty}dk\tilde{\phi}^{(n)}(v,k,\omega,\tau)\\ &\times (\omega-\lambda k)^{-1}\exp[i(k\xi-\omega\sigma)], \label{eq-F-L-int}
\end{split}
 \end{equation}
where  $W$  is the usual Laplace transform contour which excludes the poles of the integrals, and $\tilde{f}_\alpha^{(n)}$ and $\tilde{\phi}^{(n)}$ are  analytic for all $\omega$.
Equation \eqref{eq-F-L-int} represents a plane wave propagating along the $\xi$-direction with the phase speed $\lambda$.  
\par 
 Substituting the stretched coordinates \eqref{eq-stretch} and   the expansions \eqref{eq-expan1} and  \eqref{eq-F-L-int} into Eqs. \eqref{eq-wigner} and \eqref{eq-poisson} we obtain in one-dimensional geometry
 \begin{widetext}
\begin{eqnarray}
&&\left(\epsilon^{3/2} \frac{\partial}{\partial \tau}+\epsilon^{1/2}\frac{\partial}{\partial \sigma} \right)\left(f_\alpha^{(0)} +\epsilon^{1/2}f_\alpha^{(1)}+\epsilon^2 f_\alpha^{(2)}+......  \right)+\epsilon^{1/2}v\frac{\partial}{\partial \xi} \left(f_\alpha^{(0)} +\epsilon^{1/2}f_\alpha^{(1)}+\epsilon^2 f_\alpha^{(2)}+......  \right)\notag\\
&&  +\frac{em}{2i\pi \hbar^2}\int\int dr'dv' e^{im(v-v') r'/\hbar} \left[\left\lbrace \epsilon\phi^{(1)}\left(\xi+\frac{ r'}{2},\sigma,\tau\right) +\epsilon^{3/2}\phi^{(2)}\left(\xi+\frac{r'}{2},\sigma,\tau \right)+.....\right\rbrace \right. \notag\\
&& \left. -\left\lbrace\epsilon\phi^{(1)}\left(\xi-\frac{ r'}{2},\sigma,\tau\right) +\epsilon^{3/2} \phi^{(2)}\left(\xi-\frac{r'}{2},\sigma,\tau \right)+..... \right\rbrace \right]\left(f_\alpha^{(0)} +\epsilon^{1/2}f_\alpha^{(1)}+\epsilon^2 f_\alpha^{(2)}+......  \right) \doteq0, \label{wignermoyal1}
\end{eqnarray}
and
 \begin{equation}
-\epsilon \frac{\partial^2}{\partial \xi^2}\left(\epsilon \phi^{(1)}+\epsilon^{3/2} \phi^{(2)}+...... \right)= 4\pi e \sum_{\alpha=l,h}\int \left\lbrace f_\alpha^{(0)} +\epsilon^{1/2}f_\alpha^{(1)}+\epsilon^2 f_\alpha^{(2)}+......  \right\rbrace dv,  \label{poisson1}
\end{equation}
\end{widetext}
where the symbol $\doteq$ is used to denote the equality in the weak sense.  In the following subsections \ref{sec-sub-1st}-\ref{sec-sub-kdv}, we obtain different expressions for the first and second order perturbations to obtain the modified KdV equation, as well as   the conditions for  multi-plasmon resonances. 
\subsubsection{First order perturbations} \label{sec-sub-1st}
Equating the coefficients of $\epsilon$, we get from Eq. \eqref{wignermoyal1}
\begin{equation}
\begin{split}
 &\frac{\partial f_\alpha^{(1)}}{\partial \sigma}+v \frac{\partial f_\alpha^{(1)}}{\partial \xi} +\frac{e m}{2i\pi \hbar^2} \int \int dr' dv' e^{im(v-v') r'/\hbar}\\
& \times \left[\phi^{(1)}\left(\xi+\frac{\lambda}{2} \right)-\phi^{(1)}\left(\xi-\frac{\lambda}{2} \right)\right] f_\alpha^{(0)}=0. \label{eq-f1}
 \end{split}
\end{equation}
By applying to Eq. \eqref{eq-f1} the Fourier-Laplace transform, given by   Eq. \eqref{eq-F-L-int},     we obtain  
\begin{equation}
f_\alpha^{(1)}=-\frac{e}{\hbar}\frac{f_\alpha^{(0)}\left(v+\frac{\hbar k}{2m} \right) -f_\alpha^{(0)}\left(v-\frac{\hbar k}{2m} \right) }{\omega-kv} \phi^{(1)}. \label{f1-eq}
\end{equation}
In the lowest order of $\epsilon$, i.e., $\epsilon^{1/2}$, the Poisson equation \eqref{poisson1}  gives
\begin{equation}
4\pi e \sum_{\alpha=l,h} \int f_\alpha^{(1)}dv=0. \label{f1-poisson}
\end{equation}
From Eqs. \eqref{f1-eq} and \eqref{f1-poisson} eliminating $f_\alpha^{(1)}$, we obtain   the following expression for the wave speed $\lambda$.
\begin{equation}
\sum_{\alpha=l,h}\frac{\omega_p^2}{n_{\alpha_0}k^2}\int_\text{n.r.} \frac{f_\alpha^{(0)}(v)}{(\lambda-v)^2-{\hbar^2 k^2}/{4m^2}}dv=0, \label{dispersion-modified}  
\end{equation}
where   `n.r.'  stands for the non-resonance region. Equation \eqref{dispersion-modified} resembles the quasineutrality condition which agrees with the linear dispersion equation \eqref{eq-Dr} except the term `$1$' which appears as the effect of dispersion due to the charge separation of electrons. Thus, from Eq. \eqref{dispersion-modified}  it follows that   such a dispersion is a higher-order effect [at least $o(\epsilon^2)$]. 
 Note that in Eq. (\ref{dispersion-modified}) we only integrate over the non-resonant region, and thus, in order to focus on the dispersion, the small linear damping associated with the standard Landau resonance is not included here.
When the linear pole is located far enough out in the tail, the linear damping will be effectively suppressed, and such an approximation is accurate. However, as we have hinted above, this does not mean that wave-particle interaction is negligible. Importantly, as already pointed out, the nonlinear resonances can be located deeper into the background
distribution. Accordingly, even if such resonances are of higher order in the nonlinear expansion parameter, such terms do not suffer the same type of exponential suppression (due to a lower resonant velocity), and hence they
can be important even when the linear damping is small or negligible. 
 The effects of Landau damping due to the   multi-plasmon resonances will also be shown to be at least $o(\epsilon^2)$.    
\subsubsection{Second order perturbations} \label{sec-sub-2nd}
Equating the coefficients of $\epsilon^{5/2}$, we  obtain   from Eq. \eqref{wignermoyal1} 
 \begin{equation}
\begin{split}
&\frac{\partial f_\alpha^{(2)}}{\partial \sigma}+v \frac{\partial f_\alpha^{(2)}}{\partial \xi} +\frac{e m}{2i\pi \hbar^2} \int \int dr' dv' e^{im(v-v') r'/\hbar}\\
&\times\left[\phi^{(2)}\left(\xi+\frac{\lambda}{2} \right) -\phi^{(2)}\left(\xi-\frac{\lambda}{2} \right)\right]f_\alpha^{(0)}=0. \label{eq-f2}
\end{split}
\end{equation}
Transforming Eq. \eqref{eq-f2}  in view of Eq. \eqref{eq-F-L-int} and  eliminating $f^{(1)}$ by Eq. \eqref{f1-eq}, we obtain
\begin{equation}
\hat{f_\alpha}^{(2)}=-\frac{(e/\hbar)}{\omega-kv} \left[f_\alpha^{(0)}\left( v+\frac{\hbar k}{2m}\right) -f_\alpha^{(0)}\left( v-\frac{\hbar k}{2m}\right) \right]\hat{\phi}^{(2)}.\label{f2-eq}
\end{equation}
Next, from Eq. \eqref{poisson1}   equating the coefficients of $\epsilon^2$,  we get
  \begin{equation}
 \frac{\partial^2 \phi^{(1)}}{\partial \xi^2}=-4 \pi e \sum_{\alpha=l,h} \int_{-\infty}^{\infty} f^{(2)}_{\alpha} dv. \label{poisson3}
 \end{equation}
 Taking the inverse Fourier-Laplace transform of Eq. \eqref{f2-eq} and integrating with respect to $v$ over the entire velocity space, we find that the second order density perturbation  reduces  to 
\begin{equation}
\begin{split}
&\int_\text{n.r} f_\alpha^{(2)}dv=\frac{e}{\hbar k}\phi^{(2)}(\zeta,\tau)\int_\text{n.r.}(v-\lambda)^{-1}\\
&\times \left[f_\alpha^{(0)}\left(v+\frac{\hbar k}{2m} \right) -f_\alpha^{(0)}\left(v-\frac{\hbar k}{2m} \right)\right] dv,\label{modified-f2-eq} 
\end{split}
 \end{equation}
where we have defined a new coordinate $\zeta=\xi-\lambda\sigma$.  Some important points are to be noticed.    The velocity integral in Eq. \eqref{modified-f2-eq} can be shown to vanish by the dispersion equation \eqref{dispersion-modified}. Furthermore,  Eq. \eqref{f2-eq} neither involves the time evolution term nor any other nonlinear terms that involve the first-order perturbations.      So, in order to properly include a time evolution term along with the resonance and nonresonance contributions   in the nonlinear regime, a slightly different ordering  of both $f$ and $\phi$ is necessary. Thus,  we consider
 \begin{equation}
\begin{split}
f_\alpha(x,v,t)= &f_\alpha^{(0)} +\epsilon f_\alpha^{(1)}+\epsilon^2 f_\alpha^{(2)}+\cdots, \\
\phi(x,t)=&\epsilon ^{3/2} \phi^{(1)}+\epsilon^{2} \phi^{(2)}+\cdots  
\end{split} \label{expansion-resonance}
\end{equation}
 Using the expansion \eqref{expansion-resonance} and equating the coefficients of $\epsilon^{5/2}$   from Eq. \eqref{wignermoyal1}, we  obtain 
\begin{equation}
\begin{split}
\frac{\partial f_\alpha^{(2)}}{\partial \sigma}+&v \frac{\partial f_\alpha^{(2)}}{\partial \xi}+ \frac{\partial f_\alpha^{(1)}}{\partial \tau} +\frac{e m}{2i\pi \hbar^2} \int \int dr' dv' e^{im(v-v') r'/\hbar}\\
&\times\left[\phi^{(1)}\left(\xi+\frac{r'}{2} \right)f_\alpha^{(1)} -\phi^{(1)}\left(\xi-\frac{r'}{2} \right)f_\alpha^{(1)}  \right] =0. \label{f2-resonance-eq}
\end{split}
\end{equation}
By applying the Fourier-Laplace transform to Eq. \eqref{f2-resonance-eq} and eliminating  $f^{(1)}$ by Eq. \eqref{f1-eq}, we have 
\begin{equation}
\begin{split}
\hat{f_\alpha}^{(2)}=&\frac{1}{\omega-kv} \left[-\frac{ie}{\hbar \left(\omega-kv\right)} \left\lbrace f_\alpha^{(0)}\left( v+\frac{\hbar k}{2m}\right) \right.\right.\\
&\left.\left.-f_\alpha^{(0)}\left( v-\frac{\hbar k}{2m}\right)\right\rbrace  \frac{\partial \hat{\phi}^{(1)}}{\partial \tau} +\frac{e^2}{\hbar^2}\hat{\Psi}_\alpha I_\alpha\right],\label{f2-resonance-eq-fourier}
\end{split}
\end{equation}
where $\hat{\Psi}_\alpha$ is the Fourier-Laplace transform of $\left\lbrace \phi^{(1)} \right\rbrace^2$, given by,
\begin{equation}
\hat{\Psi}_\alpha=i\int_{-\infty}^{\infty}d\xi\int_{-\infty}^{\infty}d\sigma(\omega-\lambda k)\left\lbrace \phi^{(1)} (\xi,\sigma)\right\rbrace^2  e^{-i(k\xi-\omega \sigma)},  \label{B11-eq}
\end{equation}
 and 
\begin{equation}
\begin{split}
I_\alpha(v)=&\left[ \frac{f_\alpha^{(0)}\left( v+\frac{\hbar k}{m}\right)-f_\alpha^{(0)}(v) }{\omega-k\left(v+\frac{\hbar k}{2m} \right) }\right.\\
&\left.-\frac{f_\alpha^{(0)}\left( v-\frac{\hbar k}{m}\right)-f_\alpha^{(0)}(v) }{\omega-k\left(v-\frac{\hbar k}{2m} \right) }\right].\label{I(v)-eq}
\end{split}
\end{equation} 
Taking the   Fourier-Laplace inverse transform of Eq. \eqref{f2-resonance-eq-fourier} and then integrating over the velocity space, the second order density perturbation reduces  to 
\begin{equation}
\int_{-\infty}^{\infty}  f_\alpha^{(2)}dv=\frac{e}{\hbar}\left(A_\alpha+\frac{e}{\hbar} B_\alpha \right),\label{modified-f2-eq-resonance} 
\end{equation}
where 
\begin{equation}
\begin{split}
A_\alpha=&\frac{1}{(2\pi)^{2}} \int \int_W dk d\omega (\omega-\lambda k)^{-1} \frac{\partial \hat{\phi}^{(1)}}{\partial \tau}\\
&\times J_\alpha(k,\omega)\exp[i(k\xi-\omega\sigma)], \label{A-eq}
\end{split}
\end{equation}
\begin{equation}
\begin{split}
J_\alpha(k,\omega)=&\int_{-\infty}^{\infty} (\omega-kv)^{-2} \left\lbrace f_\alpha^{(0)}\left(v+\frac{\hbar k}{2m} \right) \right.\\
&\left. -f_\alpha^{(0)}\left(v-\frac{\hbar k}{2m} \right) \right\rbrace dv, \label{eq-J}
\end{split}
\end{equation}
\begin{equation}
\begin{split}
B_\alpha=&\frac{i}{(2\pi)^2}\int \int_W' dk' d\omega' (\omega'-\lambda k')^{-1}\hat{\Psi}_\alpha(\omega',k')\\
&\times L_\alpha (k,\omega) \exp[i(k'\xi-\omega'\sigma)],\label{B-eq}
\end{split}
\end{equation}
\begin{equation}
L_\alpha (k,\omega)=\int_{-\infty}^{\infty} I_\alpha(v)(\omega-kv)^{-1}dv.\label{L-eq}
\end{equation}
\subsubsection*{Multi-plasmon resonance} \label{sec-sub-reso}
Inspecting on the second-order perturbations we find that  Eq. \eqref{eq-J} can be rewritten as
\begin{equation}
\begin{split}
J_\alpha(k,\omega)=&\frac{1}{k^2}\int_{-\infty}^{\infty} \left[\frac{1}{(v-\omega/k-v_q)^2}\right.\\
&\left.- \frac{1}{(v-\omega/k+v_q)^2} \right]f_\alpha^{(0)}  dv, \label{eq-J1}
\end{split}
\end{equation}
Also,  from Eqs. \eqref{I(v)-eq} and \eqref{L-eq} we have
 \begin{equation}
 \begin{split}
 &L_\alpha(k,\omega)= \frac{1}{k^2v_q}\int_{-\infty}^{\infty} \left(\frac{1}{v-\omega/k-2v_q} \right.\\
 &\left.+\frac{1}{v-\omega/k+2v_q}-\frac{2}{v-\omega/k}\right) f_\alpha^{(0)} (v)dv. \label{eq-reso}
 \end{split}
\end{equation} 
Thus, looking at the denominators of the integrands in Eqs. \eqref{eq-J1} and \eqref{eq-reso}, we note that there are poles  at $v=\omega/k\pm n v_q$, where $n=0,1,2$. This implies that  in the propagation of nonlinear EAWs,   the phase velocity resonance,  as well as the one- (for $n=1$) and the two-plasmon (for $n=2$) resonances can  occur  in  the wave-particle interactions.  However,   the resonant velocities  $v=\omega/k+ n v_q$, $n=1,2$ are not of interest to the present study  as the lower resonant velocity $v=\omega/k-nv_q$ (for $n=0,1,2$) is more likely to cause the necessary wave damping. 
\par 
In what follows, we calculate the velocity integrals   $J_\alpha$ and $L_\alpha$   by considering $\int_{-\infty}^{\infty}dv=\int_\text{res.}dv+\int_\text{n.r.}dv$ and $\int_\text{n.r.}dv\approx {\cal P}\int_{-\infty}^{\infty}dv$, and choose the Landau contour  encircling the pole $v=\omega/k-n v_q$ below the real axis  for $\Im\omega<0$. Then the integration with respect to  $\omega$ can be performed by closing the $W$ contour in the lower-half plane encircling the pole at $\omega=\lambda k$. Thus, for $r=0,1,2,...$ and $n=1,2$ we obtain
\begin{equation}
\begin{split}
&\lim_{\omega\rightarrow{\lambda k}} \int_{-\infty}^{\infty} \frac{f_\alpha^{(0)}}{\left[v-(\omega/k-n v_q)\right]^{r+1}}dv\\
&=  {\cal P} \int_{-\infty}^{\infty} \frac{f_\alpha^{(0)}}{\left[v-(\lambda-n v_q)\right]^{r+1}}dv \\
&+i\pi \frac{k}{|k|} \frac{1}{r!} \left[\frac{d^r}{dv^r} f_\alpha^{(0)}(v)\right]_{v=\lambda-n v_q},
\end{split}
\end{equation}
\begin{equation}
\begin{split}
&\lim_{\omega\rightarrow{\lambda k}}J_\alpha(k,\omega)\equiv J_\alpha(k,\lambda) =J_{1\alpha}(k,\lambda)+iJ_{2\alpha}(k,\lambda),\\
&\lim_{\omega\rightarrow{\lambda k}}L_\alpha(k,\omega)\equiv L_\alpha(k,\lambda) =L_{1\alpha}(k,\lambda)+iL_{2\alpha}(k,\lambda),
\end{split}
\end{equation}
where  
\begin{equation}
J_{1\alpha}(k,\lambda)=\frac{1}{k^2}\int_C \left[\frac{1}{(v-\lambda-v_q)^2}- \frac{1}{(v-\lambda+v_q)^2} \right]f_\alpha^{(0)}  dv, \label{eq-J11}
\end{equation}
\begin{equation}
J_{2\alpha}(k,\lambda)= -\frac{\pi}{k^2}\frac{k}{|k|}\left[ \frac{d}{dv} f_\alpha^{(0)}(v)\right]_{v=\lambda-v_q}, \label{L2-eq} 
\end{equation}
 \begin{equation}  \label{L1-eq}
 \begin{split}
 L_{1\alpha}(k,\omega)= &\frac{1}{k^2v_q}\int_C \left(\frac{1}{v-\lambda-2v_q}+\frac{1}{v-\lambda+2v_q}\right.\\
 &\left.-\frac{2}{v-\lambda}\right) f_\alpha^{(0)} (v)dv,  
 \end{split}
\end{equation} 
\begin{equation}
L_{2\alpha}= \frac{\pi}{k^2v_q} \frac{k}{|k|} \left[f_\alpha^{(0)} \left(v-\lambda+2v_q\right)-2f_\alpha^{(0)}\left(v-\lambda\right)\right]. \label{L2-eq} 
\end{equation}
Here, $C$ is the Landau contour which excludes the poles at $v=\lambda\pm n v_q$, $n=0,2$.
\subsubsection*{The KdV equation} \label{sec-sub-kdv}
Noting  that 
\begin{equation}
\begin{split}
&\frac{i}{(2\pi)^2}\int dk \int d\omega (\omega-\lambda k)^{-1} e^{i(k\xi-\omega\sigma)}\frac{k}{|k|} \hat{\Psi}_\alpha(k,\omega)\\
&=-i\frac{\cal P}{\pi} \int d\zeta'\frac{\left\lbrace \phi^{(1)}(\zeta,\lambda) \right\rbrace^2 }{\zeta-\zeta'},\label{P-eq}
\end{split}
\end{equation}
 we finally obtain from Eq. \eqref{B-eq}
\begin{equation}
B_\alpha=\left\lbrace\phi^{(1)} \right\rbrace^2 L_{1\alpha}+L_{2\alpha} \frac{\cal P}{\pi} \int d\zeta'\frac{\left\lbrace \phi^{(1)}(\zeta,\lambda) \right\rbrace^2 }{\zeta-\zeta'}.\label{modified-B-eq}
\end{equation}
Next,  differentiating $A_\alpha$ and $B_\alpha$ with respect to $\zeta=\xi-\lambda\sigma$ we get
\begin{equation}
\frac{\partial A_\alpha}{\partial \zeta}=kJ_{\alpha}(k,\lambda)\frac{\partial \phi^{(1)}}{\partial \tau}, \label{A-derivative-eq}
\end{equation} 
and
\begin{equation}
\begin{split}
\frac{\partial B_\alpha}{\partial \zeta}=&2\phi^{(1)}\frac{\partial \phi^{(1)}}{\partial \zeta}L_{1\alpha}\\
&+L_{2\alpha}\frac{\cal P}{\pi}\int \frac{1}{\zeta-\zeta'}\frac{\partial}{\partial \zeta}\left\lbrace\phi^{(1)}\left(\zeta,\tau\right)\right\rbrace^2   d\zeta'.\label{B-derivative-eq}
\end{split}
\end{equation}
In the frame $\zeta=\xi-\lambda\sigma$, Eq. \eqref{poisson3} reduces to
 \begin{equation}
 \frac{\partial^2 \phi^{(1)}}{\partial\zeta^2}=-4 \pi e \sum_{\alpha=l,h}  \int_{-\infty}^{\infty} f^{(2)}_{\alpha} dv.\label{poisson2}
 \end{equation}
Differentiating Eq. \eqref{poisson2} once with respect to $\zeta$ and using the relations [i.e., Eqs. \eqref{modified-f2-eq},  \eqref{modified-f2-eq-resonance},  \eqref{A-derivative-eq} and \eqref{B-derivative-eq}],  we obtain the following modified KdV equation with nonlinear Landau damping $\propto\Gamma$.
\begin{equation}
\frac{\partial \phi}{\partial \tau}+A\frac{\partial ^3\phi}{\partial \zeta^3}+B \phi\frac{\partial \phi}{\partial \zeta} +  \Gamma{\cal P}\int_{-\infty}^{\infty} \frac{\partial \phi^2\left(\zeta',\tau\right)}{\partial \zeta'} \frac{1}{\zeta-\zeta'}    d\zeta'=0,\label{K-dV1}
\end{equation}
where $\phi\equiv\phi^{(1)}$, ${\cal P}$ denotes the Cauchy principal value, and the coefficients of the KdV equation are $A=1/P$, $B=Q/P$ and $\Gamma=R/P$ with
\begin{equation}
\begin{split}
P=&\frac{4\pi e^2}{\hbar k} \sum_{\alpha=l,h}\int_C \left[\frac{1}{(v-\lambda-v_q)^2}- \frac{1}{(v-\lambda+v_q)^2} \right]f_\alpha^{(0)}  dv\\
&+i\sgn (k)\frac{4\pi^2 e^2}{\hbar k }\sum_{\alpha=l,h}\left[ \frac{d}{dv} f_\alpha^{(0)}(v)\right]_{v=\lambda-v_q}, \label{alpha-eq}
\end{split}
\end{equation}
\begin{equation}
\begin{split}
Q=&\frac{16\pi me^3}{(\hbar k)^3} \sum_{\alpha=l,h}\int_C \left(\frac{1}{v-\lambda-2v_q}\right.\\
&\left.+\frac{1}{v-\lambda+2v_q}-\frac{2}{v-\lambda}\right) f_\alpha^{(0)} (v)dv \label{beta-eq}
\end{split}
\end{equation}
and
\begin{equation}
R=\frac{8\pi me^3}{(\hbar  k)^3} \sum_{\alpha=l,h}\left[ f_\alpha^{(0)}\left(v-\lambda+2v_q\right)-2f_\alpha^{(0)}\left(v-\lambda\right)\right]. \label{gamma-eq}
\end{equation}
From the coefficients of the KdV equation \eqref{K-dV1} we note that $P$ becomes complex due to the one plasmon resonance, and so are the dispersive $(\propto A)$, local nonlinear $(\propto B)$, and  the nonlocal nonlinear  $(\propto\Gamma)$ terms. It is interesting to note that although the form of the KdV equation \eqref{K-dV1} looks similar to that first obtained by Ott and Sudan \cite{ott1969} and later by many authors (See, e.g., Refs. \citep{vandam1973,barman2017}) in classical plasmas, however,      the   Landau damping term $\propto\Gamma$ appears here as nonlinear    due to the  phase velocity resonance as well as the two-plasmon  resonance processes in the wave-particle interactions. The appearance of such a nonlocal nonlinearity not only modifies the propagation of EAWs but also introduces a new wave damping mechanism which has not been studied before in the context of KdV theory of small-amplitude waves.            
  \par 
To get further insight of the KdV equation,  we simplify the coefficients $P$, $Q$ and $R$ by assuming the smallness in the argument of the logarithmic function in Eq. \eqref{eq-distb-fn}. This is valid since $\xi_\alpha\equiv \mu_\alpha/kBT_\alpha$ may be large and negative for $T_\alpha>T_{F\alpha}$. Also, we evaluate  the integrals by means of Eq. \eqref{eq-int} and   approximate them as  
\begin{equation}
\begin{split}
&{\cal P} \int _{-\infty}^{\infty}\frac{\exp(-v^2/v^2_{t\alpha})}{v-\lambda\pm v_q}dv\\
&=\left\lbrace
\begin{array}{cc}
\frac{4}{3}-2\sqrt{\pi}\delta_{\alpha\pm}-\frac{14}{15}\delta^2_{\alpha\pm}+\cdots, & \delta_{\alpha\pm}<1   \\
   -\frac{1.8}{\delta_{\alpha\pm}}-\frac{0.89}{\delta^3_{\alpha\pm}}-\cdots, & \delta_{\alpha\pm}>1  
\end{array}\right., 
\end{split} 
\end{equation}
where  $\delta_{\alpha\pm}=(\lambda\pm v_q)/\sqrt{2}v_{t\alpha}$.
In this way,  the expressions for $P$, $Q$ and $R$ reduce to
  \begin{equation}
  \begin{split}
  &\Re P=-\frac{6\pi e^2}{m}\sum_{\alpha=l,h}\frac{n_{\alpha0}T_\alpha e^{\xi_\alpha}}{v_{F\alpha}T_{F\alpha}}\left[\frac{5(\lambda^2-v_q^2)+2\lambda v_{t\alpha}}{(\lambda^2-v_q^2)^2}\right],
  \end{split}
  \end{equation}
  \begin{equation}
  \begin{split}
 \Im P=&-6(\lambda-v_q)\frac{\pi^2 e^2}{\hbar k}\sum_{\alpha=l,h}\frac{n_{\alpha0}T_\alpha e^{\xi_\alpha}}{v_{F\alpha}T_{F\alpha}v_{t\alpha}^2}\\
  &\times\exp\left\lbrace-\left(\frac{\lambda-v_q}{v_{t\alpha}}\right)^2\right\rbrace,
  \end{split}
  \end{equation}
  \begin{equation}
  Q=-\frac{24\pi me^3}{(\hbar k)^3}\frac{\lambda^2+4v_q^2}{\lambda(\lambda^2-4v_q^2)}\sum_{\alpha=l,h}\frac{n_{\alpha0}T_\alpha v_{t\alpha}}{v_{F\alpha}T_{F\alpha}}e^{\xi_\alpha},
  \end{equation}
  \begin{equation}
  \begin{split}
  R=\frac{6\pi^2m e^3}{(\hbar k)^3}&\sum_{\alpha=l,h}\frac{n_{\alpha0}T_\alpha }{v_{F\alpha}T_{F\alpha}}e^{\xi_\alpha}\left[\exp\left\lbrace-\left(\frac{\lambda-2v_q}{v_{t\alpha}}\right)^2\right\rbrace\right.\\
 &\left. -2\exp\left\lbrace-\left(\frac{\lambda}{v_{t\alpha}}\right)^2\right\rbrace\right]. \label{eq-R-reduced}
  \end{split}
  \end{equation}
  From the reduced expression of $R$ [Eq. \eqref{eq-R-reduced}]  it is evident that the contribution of the two-plasmon resonance is higher than that of the phase velocity resonance, implying that the two-plasmon resonance process is the dominant wave damping mechanism for EAWs in a plasma with finite temperature degeneracy.  It is to be noted that the contribution of the two-plasmon resonance process in the Landau damping will be smaller or larger than those of the phase velocity and linear plasmon resonances depending on   how the background particles are distributed over the entire velocity space.  At finite temperature, the projected Fermi-Dirac distribution \eqref{eq-distb-fn} assumes values for all velocities less than the Fermi velocity even though it decays quickly. So damping exists for all wave numbers.  Furthermore,  since the two-plasmon  resonant velocity  $\omega/k-2v_q$ is smaller than those due to the linear Plasmon    $(\omega/k-v_q)$ and  the phase velocity $(\omega/k)$, the     two-plasmon resonance process will give the wave damping more easily. For the distribution \eqref{eq-distb-fn}, the resonance contribution due to two-plasmon process is higher than those associated with the linear Plasmon and phase velocity resonances as there are less number of particles near the tails of the distribution.  The results may be true for some other distributions, e.g., Fermi-Dirac distribution at zero temperature \cite{misra2017}. 
 \section{Conservation laws} \label{sec-conser}
It is imperative to check whether some conservation laws hold for the nonlocal KdV equation \eqref{K-dV1}. Since we look for a solitary wave solution of Eq. \eqref{K-dV1}, it may be assumed that the function $\phi(\zeta,\tau)$ and its  derivatives with respect to $\zeta$ vanish as $\zeta\rightarrow\pm\infty$. So, integrating Eq. \eqref{K-dV1} with respect to $\zeta$, we obtain
\begin{equation}
\begin{split}
 &\frac{\partial }{\partial \tau}\int_{-\infty}^{\infty}\phi(\zeta,\tau)d\zeta+A\int_{-\infty}^{\infty}\frac{\partial ^3 \phi}{\partial \zeta^3}d\zeta+\frac{B}{2}\int_{-\infty}^{\infty} \frac{\partial \phi^2}{\partial \zeta}d\zeta \\
 &+  \Gamma\int_{-\infty}^{\infty}\int_{-\infty}^{\infty}\frac{\partial \phi^2\left(\zeta',\tau\right)}{\partial \zeta'} {\cal P}\frac{1}{\zeta-\zeta'}d\zeta d\zeta'=0.\label{eq-conser1}
 \end{split}
\end{equation}
The second and third integrals in Eq. \eqref{eq-conser1} can be made zero using the above boundary conditions. Also,   the fourth integral,  after changing the order of integration,  becomes 
\begin{equation}
\begin{split}
&\int_{-\infty}^{\infty}\int_{-\infty}^{\infty}\frac{\partial \phi^2\left(\zeta',\tau\right)}{\partial \zeta'} {\cal P}\frac{1}{\zeta-\zeta'}d\zeta d\zeta'\\
&=\int_{-\infty}^{\infty}\frac{\partial \phi^2\left(\zeta',\tau\right)}{\partial \zeta'}d\zeta'\int_{-\infty}^{\infty}{\cal P}\frac{1}{\zeta-\zeta'}d(\zeta-\zeta'),
\end{split}
\end{equation}
and noting that the integral involving ${\cal P}$   vanishes identically. 
Thus, we have  
\begin{equation}
 \frac{\partial }{\partial \tau}\int_{-\infty}^{\infty}\phi(\zeta,\tau)d\zeta=0. \label{eq-conser-part}
 \end{equation}
  From the linear theory (Sec. \ref{sec-linear}) after Fourier analyzing  the Poisson equation \eqref{eq-poisson} one obtains   $-k^2\phi^{(1)}=4\pi e\sum_{\alpha=l,h} \int f_\alpha^{(1)}dv= 4\pi e\sum_{\alpha=l,h} n_{\alpha1}=4\pi e(n-n_0)$, where $\sum_{\alpha=l,h} n_\alpha=n$ is the total number density of particles.  Since $n_0$ is independent on time, the vanishing of the time derivative of $\phi\equiv\phi^{(1)}$ in Eq. \eqref{eq-conser-part} gives  that of the total number density $n$. Thus  from Eq. \eqref{eq-conser-part}, it follows that the KdV equation \eqref{K-dV1} conserves the total number of particles. 
 \par 
 Next, we multiply Eq. \eqref{K-dV1} by $\phi^*(\zeta,\tau)$ and the conjugate of Eq. \eqref{K-dV1} by $\phi(\zeta,\tau)$ successively,  and add the resulting equations. After  integration  with respect to $\zeta$  yields
  \begin{equation}
  \begin{split}
 &\frac{\partial }{\partial \tau}\int_{-\infty}^{\infty}|\phi(\zeta,\tau)|^2d\zeta+\int_{-\infty}^{\infty}\left(A\phi^*\frac{\partial ^3 \phi}{\partial \zeta^3}+A^*\phi\frac{\partial ^3 \phi^*}{\partial \zeta^3}\right)d\zeta\\
 &+\int_{-\infty}^{\infty}\left(B |\phi|^2\frac{\partial \phi}{\partial \zeta}+B^* |\phi|^2\frac{\partial \phi^*}{\partial \zeta}\right)d\zeta \\ 
 &+  \int_{-\infty}^{\infty}\int_{-\infty}^{\infty}\left( \Gamma \phi^*(\zeta,\tau)\frac{\partial \phi^2\left(\zeta',\tau\right)}{\partial \zeta'} {\cal P}\frac{1}{\zeta-\zeta'}\right.\\
 &\left.+ \Gamma^*  \phi(\zeta,\tau)\frac{\partial \phi^{*2}\left(\zeta',\tau\right)}{\partial \zeta'} {\cal P}\frac{1}{\zeta-\zeta'}\right)d\zeta d\zeta'=0.\label{eq-conser2}
 \end{split}
\end{equation}
The second and third pairs of integrals of Eq. \eqref{eq-conser2} vanish by evaluating the integrals by parts and using the boundary conditions stated above. The fourth pair of integrals of Eq. \eqref{eq-conser2} can be evaluated by using the fact that $\hat{\phi^*}(k,\tau)=\hat{\phi}^*(-k,\tau)$,  $\hat{\phi}(k,\tau)=-2ik|\hat{\phi}(k,\tau)|^2$, $k\sgn (k)=|k|$, and 
\begin{equation}
  \int_{-\infty}^{\infty} \phi^*\left(\zeta,\tau\right) {\cal P}\frac{1}{\zeta-\zeta'} d\zeta= i\pi\int_{-\infty}^{\infty}\sgn(k)\hat{\phi^*}(k,\tau)e^{ik\zeta'}dk. 
\end{equation}
So, we obtain    
\begin{equation}
\begin{split}
  &\int_{-\infty}^{\infty}\int_{-\infty}^{\infty}\phi^*(\zeta,\tau)\frac{\partial \phi^2\left(\zeta',\tau\right)}{\partial \zeta'} {\cal P}\frac{1}{\zeta-\zeta'}\\
  &=\int_{-\infty}^{\infty}\int_{-\infty}^{\infty}\phi(\zeta,\tau)\frac{\partial \phi^{*2}\left(\zeta',\tau\right)}{\partial \zeta'} {\cal P}\frac{1}{\zeta-\zeta'}d\zeta d\zeta'\\
    &=-4\pi\int_{-\infty}^{\infty}k^2|k||\hat{\phi}(k,\tau)|^2|\hat{\phi^2}(k,\tau)|^2dk<0.
    \end{split}
  \end{equation}
 Since $\Gamma=R/P$ with $R>0$ and $\Re P<0$, we have $\Gamma+\Gamma^*<0$. Then Eq. \eqref{eq-conser2} reduces to 
 \begin{equation}
 \begin{split}
&\frac{\partial }{\partial \tau}\int_{-\infty}^{\infty}|\phi(\zeta,\tau)|^2d\zeta\\
&+\tilde{\Gamma}\int_{-\infty}^{\infty}k^2|k||\hat{\phi}(k,\tau)|^2|\hat{\phi^2}(k,\tau)|^2dk=0, \label{eq-conser-redu}
\end{split}
\end{equation} 
 where $\tilde{\Gamma}=-4\pi(\Gamma+\Gamma^*)>0$. The first term in Eq. \eqref{eq-conser-redu} represents the rate of change of  wave energy whereas the second one is the rate of change of energy of resonant particles. Thus, the total energy is conserved. However, since the second term is positive definite, by the $H$-theorem we have 
 \begin{equation}
\frac{\partial }{\partial \tau}\int_{-\infty}^{\infty}|\phi(\zeta,\tau)|^2d\zeta\leq0, \label{eq-cons-enr}
\end{equation} 
where the equality holds at $\phi=0~ \forall~ \zeta$. This means that the wave energy decays with time and hence a steady-state solution with finite wave energy does not exist.
\section{Solitary wave solution} \label{sec-soliton}
We note that in absence of the damping effect, i.e., $\Gamma=0$, Eq. \eqref{K-dV1} reduces to the usual KdV equation:
\begin{equation}
\frac{\partial \phi}{\partial\tau}+A \frac{\partial^3 \phi}{\partial \zeta^3}+B \phi\frac{\partial\phi}{\partial\zeta}=0. \label{K-dv-usual}
\end{equation}
A traveling solitary wave solution of Eq. \eqref{K-dv-usual} is known, i.e., 
\begin{equation}
\phi=\phi_0 \rm{sech}^2 \left( \frac{\zeta-U_0\tau}{\cal W}\right),  \label{eq-sol-solit}
\end{equation}
where  $\phi_0={3U_0}/{B}$ is the constant amplitude,    ${\cal W}=\sqrt{12A/\phi_0B } =\sqrt{4A/U_0}$ is the constant width, and   $U_0=\phi_0 B/3$ is the constant phase speed of electron-acoustic solitary waves.
\par 
Our aim is to find an approximate solitary wave solution of the KdV equation \eqref{K-dV1} with a small effect of the nonlinear Landau damping $(\propto\Gamma)$ due to the phase velocity and multi-plasmon resonances. We also assume that $|A|,~|B|\gg|\Gamma|\sim\epsilon$. From Eq. \eqref{eq-cons-enr}   it is evident that an initial perturbation   of the form of Eq. \eqref{eq-sol-solit} will decay to zero with time. So, in presence of the nonlinear Landau damping   the solitary wave amplitude   $\phi_0$ is no longer a constant (but decreases with time), i.e., $\phi_0\equiv\phi_0(\tau)$. Thus,  we introduce  a new space co-ordinate $z$ in a frame moving with the solitary wave and normalized to its width as
 \begin{equation}
 z=\left(\zeta-\frac{1}{3}B\int_0^\tau \phi_0(\tau) d\tau\right)\Big/ {\cal W}, \label{space-coordinate}
 \end{equation}
 and  two time scales, namely   $\eta_0=\tau$ and $\eta_1=\Gamma\tau$. Next, we expand the solution   with the scaling parameter $\Gamma\sim\epsilon$  as
 \begin{equation}
 \phi(z,\tau)=\phi^{(0)}+\Gamma \phi^{(1)}+\Gamma^2\phi^{(2)}+...,\label{solution}
 \end{equation}
where $\phi^{(i)}$ is a function  of $\tau=\tau_0,~\tau_1,~ \tau_2,...$ with $\tau_i=\Gamma^i\tau$ for $i=0,1,2,3...$
Substituting Eqs. \eqref{space-coordinate} and \eqref{solution} into Eq. \eqref{K-dV1},  we get
\begin{equation}
\begin{split}
 &\left[\rho\frac{\partial \phi^{(0)}}{\partial \eta_0}+\frac{\partial^3\phi^{(0)}}{\partial z^3}+4\left(3\frac{\phi^{(0)}}{\phi_0}-1 \right)\frac{\partial \phi^{(0)}}{\partial z}  \right] \\
 &+\Gamma\left\lbrace\rho\frac{\partial \phi^{(1)}}{\partial \eta_0}+\frac{\partial}{\partial z}\left[ \frac{\partial^2\phi^{(1)}}{\partial z^2}+4\left(3\frac{\phi^{(0)}}{\phi_0}-1 \right) \phi^{(1)}\right]\right.\\
 &\left. -\rho M[\phi^{(0)}]  \right\rbrace =0(\Gamma^2), \label{equation}
 \end{split}
\end{equation}
 where 
 \begin{equation}
 \rho=\sqrt{A}\left(\frac{1}{12}\phi_0 B \right)^{-3/2}, \label{rho-eq} 
 \end{equation}
  \begin{equation}
  \begin{split}
 &M[\phi^{(0)}]\equiv-\left[\frac{\partial \phi^{(0)}}{\partial \eta_1}+\frac{z}{2\phi_0(\tau)}\frac{\partial \phi_0(\tau)}{\partial\eta_1}\frac{\partial \phi^{(0)}}{\partial z}\right.\\
 &\left.+\sqrt{\frac{\phi^{(0)}B}{12A}} {\cal P} \int_{-\infty}^{\infty} \frac{\partial}{\partial z}\left\lbrace \phi^{(0)}\left(z,\tau\right)\right\rbrace ^2  \frac{dz'}{z-z'}\right].\label{M-eq}
 \end{split}
 \end{equation}
Applying the  initial and boundary conditions, i.e., 
 \begin{equation}
   \phi(z,0)=\phi_{00}\text{sech}^2 z,~ 
  \phi(\pm\infty,\tau)=0, \label{ini-boun-cond}
 \end{equation}
 where $\phi_{00}=\phi_0(\tau=0)$ and solving the order of unity equation we get
\begin{equation}
\phi^{(0)}(z,\eta_0,\eta_1)=\phi_0(\eta_1)\text{sech}^2z, \label{unity-eq}
\end{equation}
Next, to the order of $\Gamma$, we have
\begin{equation}
\frac{\partial \phi^{(1)}}{\partial \eta_0}+L[\phi^{(1)}]=M[\phi^{(0)}], \label{1st-order-eq}
\end{equation}
with  $\phi^{(1)}(\pm\infty,\eta_0)=0$,   $\phi^{(1)}(\beta,0)=0$, and  
 \begin{equation}
 L[\phi^{(1)}]=\frac{1}{\rho}\frac{\partial}{\partial z}\left[\frac{\partial^2}{\partial z^2}+4(3\text{sech}^2z-1) \right]\phi^{(1)}. \label{L-phi-eq} 
\end{equation}
For the existence of solutions of Eq. \eqref{1st-order-eq}, it is necessary that $M[\phi^{(0)}]$ must be orthogonal to all solutions $g(z)$ of $L^+(g)$ which satisfy   $g(\pm\infty)=0$, where $L^+$ is the operator adjoint to $L$, defined by,

\begin{equation}
\int_\infty^\infty \psi_1(z)L[\psi_2(z)]dz=\int_\infty^\infty \psi_2(z)L^+[\psi_1(z)]dz,
\end{equation}
 with  $\psi_1(\pm\infty)=\psi_2(\pm\infty)=0$ and the only solution of $L^+(g)=0$
is $g(z)=\text{sech}^2z$. Thus, we have
\begin{equation}
\int_\infty^\infty \text{sech}^2zM[\phi^{(0)}]dz=0,
\end{equation}
which gives
\begin{equation}
\begin{split}
&\frac{\partial\phi_0}{\partial \tau}+\frac{\Gamma}{2}\sqrt{\frac{B}{3A}}\phi_0^{5/2}\\
&\times{\cal P}\int_\infty^\infty \int_\infty^\infty \text{sech}^2z\frac{\partial}{\partial z}\left( \text{sech}^4z\right)\frac{dz dz'}{z-z'} =0. \label{final-phi-eq}
\end{split}
\end{equation}
Finally, we obtain for the KdV equation \eqref{K-dV1}
\begin{equation}
\phi_0(\tau)=\phi_{00}\left( 1+\frac{\tau}{\tau_0}\right)^{-2/3}, \label{phi-solution}  
\end{equation}
where 
\begin{equation}
\tau_0^{-1}=\frac{3}{4}\Gamma\sqrt{\frac{B}{3A}}\phi_{00}^{3/2}{\cal P}\int_\infty^\infty \int_\infty^\infty \text{sech}^2z\frac{\partial}{\partial z}\left( \text{sech}^4z\right)\frac{dz dz'}{z-z'}, \label{tau-eq}
\end{equation}
From Eq. \eqref{phi-solution} it is noted that  the nonlinear  Landau damping causes the wave amplitude to decay with time $\sim(\tau+\tau_0)^{-2/3}$. This decay rate is   slower than $[\sim(\tau+\tau_0)^{-2}$ due to linear Landau damping]  that  predicted by Ott and Sudan \cite{ott1969} in classical plasmas.
\section{Discussion and Conclusion} \label{sec-conclusion}
Using the Wigner-Moyal equation, we have studied  the linear and nonlinear wave particle-interactions and the evolution of  EAWs in a partially degenerate quantum plasma with two groups of electrons having  temperatures larger than the Fermi temperature.  We have  considered the background distributions of electrons as given by the Fermi-Dirac distribution at finite temperature. With this choice the linear resonance can not be ruled out as there  always remain   a few percentage of particles in the tail of the Fermi-Dirac distribution that can also participate  in the wave-particle interactions.     
\par 
 In the linear limit, a general (linear) dispersion relation  is derived which  manifests the coupling of high-frequency Langmuir waves (LWs) and the relatively low-frequency EAWs. While the high-frequency branch is recovered in the limit $\omega_r/k\pm v_q\gg v_{th}$, the low-frequency EAW mode is obtained in the opposite limit, i.e., $\omega_r/k\pm v_q\ll v_{tl}$. Interestingly, this EAW mode  has the requisite properties that apply to the KdV theory, and     for small wave number it has the form $\omega_r\approx 2.5kv_{th}$, i.e.,    the wave frequency is bit higher than that of the classical mode   $\omega_r\approx 1.31kv_{th}$  \cite{valentini2012}. Furthermore, in contrast to the classical mode and depending on the temperature or density ratios of the two species of electrons, the EAWs can turn over, going to zero, and then assume negative values for some values of the wave number  $K\equiv k \lambda_h$ beyond its critical value $K_c$.   However, the linear damping rate $|\gamma_L/\omega_{ph}|$ remains small $(\sim0.01)$ at some lower and higher values of $K$ within  $0\lesssim K<1$.   This enables us to  consider the regime $0\lesssim K\lesssim 0.5$ or  $\hbar k/mv_{th}\lesssim0.2$ such that the Wigner-Moyal equation is still applicable to the present theory. 
 \par 
 Using the multiple scale expansion technique, the evolution of weakly nonlinear EAWs is shown to be governed by a modified KdV equation with a nonlocal nonlinearity. The latter appears due to the phase velocity as well as the multi-plasmon resonances. Since the KdV equation accounts for upto the second order perturbations, the lower resonance velocity is that of the two-plasmon processes. We have also verified the conservation laws. It is found that while the total number of particles is conserved, the wave energy decays with time and so a steady-state solution of the KdV equation with finite wave energy does not exist.  An approximate analytic solitary wave solution of the KdV equation is also obtained.  It is shown that the wave amplitude   decays with time $\sim(\tau+\tau_0)^{-2/3}$ which    is slower than $\sim(\tau+\tau_0)^{-2}$ (due to linear Landau damping) as predicted by Ott and Sudan \cite{ott1969} in classical plasmas. 
 \par 
 In conclusion,   we have considered a plasma which is not in thermodynamic equilibrium.  The theory of EAWs can be studied for   plasmas in thermodynamic equilibrium (i.e., for a single species with a temperature), however, for such a background distribution as Eq. \eqref{eq-distb-fn} the EAWs tend to have a higher damping rate, making the nonlinear analysis less important.  The results are independent of the background distribution of electrons. One can consider any other form of the background distribution function that leads EAWs  fulfill the right properties for its description by the KdV equation. For example, Fermi-Dirac  distribution with arbitrary degeneracy of electrons. However, in the non-degenerate limit $(T_\alpha\gg T_{F\alpha})$, the Fermi-Dirac distribution reduces to that of the Maxwell-Boltzmann  for which    Vlasov equation applies. The motivation for considering the finite temperature $(T_\alpha>T_{F\alpha})$ is that the use of the Wigner-Moyal equation (rather than the Vlasov equation) brings about the appearance of   multi-plasmon resonances along with the phase velocity resonance. Also, the lower [than the phase velocity $(\omega/k)$] resonant velocities  $\omega/k-n v_q$, $n=1,2$  that are associated with the resonant particles absorbing wave quanta can give the wave damping more easily, and the magnitudes of the resonance terms with  $\omega/k-n v_q$ are rather higher than that are associated with $\omega/k$.  This means that the two-plasmon resonance process is the dominant wave-damping mechanism in the description of EAWs in quantum plasmas.   
   On the other hand, the choice of a fully degenerate $(T_\alpha\ll T_{F\alpha})$  background is also inadmissible as electrons with two groups can not be fully degenerate from  quantum mechanical points of view.  So, a more general background distribution which will include these as special cases  may be considered     for better understanding the underlying physics of wave-particle interactions in degenerate plasmas. However, this is left for a future work. 
\begin{acknowledgments}
 This work was  partially supported by the SERB (Government of India) sponsored research project with sanction order no. CRG/2018/004475. One of the authors (D. Chatterjee) acknowledges support from Science and Engineering Research Board (SERB)  for a national postdoctoral fellowship (NPDF)  with sanction order no. PDF/2020/002209 dated 31 Dec 2020.
\end{acknowledgments}
\section*{Data Availability} Data sharing is not applicable to this article as no new data were created or analyzed in this study.

\bibliographystyle{apsrev4-1}
\bibliography{Reference}

\end{document}